\documentclass{an_2c_tr}
\usepackage{psfig}
\begin{document}
\sloppy
\title{The Empirical X-Ray Luminosity -- Gravitational Mass
Relation for Clusters of Galaxies}
\author{T.h. Reiprich%\thanks{Email: reiprich@mpe.mpg.de} 
\and  H. B\"ohringer}
\institute{
Max-Planck-Institut f\"ur extraterrestrische Physik,
Giessenbachstra\ss e 1, D-85740 Garching, Germany
}
\headnote{Astron. Nachr. 320 (1999)}
\maketitle
\section{Introduction}
The ROSAT All-Sky Survey (RASS) is currently the best data base to
construct X-ray flux-limited samples of nearby X-ray clusters (e.\,g.,
Ebeling et al.\ 1998, B\"ohringer et al.\ 1998). Fully exploiting the
RASS, along with optical follow up, will yield X-ray luminosities
($L_{\rm X}$) for
$\sim 2000$ galaxy clusters in the near future. Assigning
gravitational masses ($M_{\rm tot}$) to
these clusters is of vital importance to probe cosmological models, e.\,g., by
translating the cluster luminosity function to the mass function and
comparison to numerical and analytical models (e.\,g., Press \&
Schechter 1974). Determining masses individually for this large number
of clusters is not feasible at present. Alternatively one can test
observationally if there is
a correlation between $L_{\rm X}$ and $M_{\rm tot}$ using a smaller sample. 
In the course of constructing an X-ray flux-limited sample of the
brightest galaxy clusters
to determine the cluster mass function (Reiprich 1998; Reiprich \&
B\"ohringer 1999) we derived gravitational masses for 106 
clusters. The mass estimates are based on the hydrostatic assumption and
are carried out individually for each cluster but in a homogeneous way
for the sample. We find a good correlation
of $L_{\rm X}$ and $M_{\rm tot}$. We use $H_0=50\,\rm
\frac{km}{s \cdot Mpc}$
and $q_0=0.5$
throughout.
\section{Data Reduction and Analysis}
We used mainly high exposure ROSAT PSPC pointed observations to
determine the physical parameters of the clusters. If clusters
extend the PSPC's field of
view or no pointed PSPC observation was available, RASS
data were used. We determined the intracluster gas density profile
using the isothermal $\beta$-model (Cavaliere \& Fusco-Femiano
1976). Using this gas density profile and assuming the gas to be
isothermal we calculated $M_{\rm tot}(<r)$ by the hydrostatic equation. For
the average gas temperature ($T_{\rm gas}$) we employed mainly published ASCA values, but also
values determined with previous satellites. For 24 clusters
we estimated $T_{\rm gas}$
using the 
$L_{\rm X}$ -- $T_{\rm gas}$ relation given by Markevitch (1998).
There
are currently contradictory measurements for average radial $T_{\rm
gas}$ gradients (e.\,g., Markevitch et al.\ 1998, Irwin et al.\ 1999,
Kikuchi et al.\ 1999). If
the $T_{\rm gas}$ decrement towards outside is commonly
present, the isothermal assumption leads to
a slight systematic overestimation of $M_{\rm tot}$ inside our adopted
outer radius. If
confirmed, inclusion of the averaged $T_{\rm gas}$ profile will be
performed.
To be able to compare clusters of different size we determined $M_{\rm
tot}$ inside the radius $r_{500}$ where the gravitational mass density
equals 500 times the critical density.
\section{Results}
\begin{figure}[h]
\psfig{file={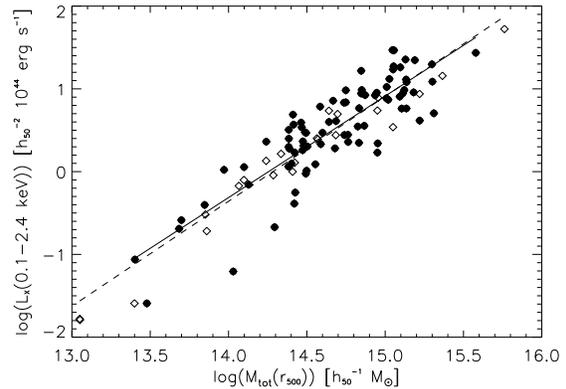},width=8.0cm,angle=0,clip=}
\caption{$L_{\rm X}$ -- $M_{\rm tot}$ relation for 106 galaxy
clusters (dashed line). Open diamonds indicate clusters where $T_{\rm gas}$ was
determined using the $L_{\rm X}$ -- $T_{\rm gas}$ relation. These
clusters were excluded for the relation given in section
3 (solid line).}\label{1}
\end{figure}

In Fig.\ \ref{1} the resulting $L_{\rm X}$ -- $M_{\rm tot}$ relation is
shown. $L_{\rm X}$ is determined in the ROSAT energy band 0.1 -- 2.4
keV. The lines represent linear regression fits to the
logarithmically plotted data points. The preliminary result is
$L_{\rm X}=3.3\cdot 10^{26}\, M_{\rm tot}^{1.23}(<r_{500})$; $L_{\rm X}$ in
erg/s and $M_{\rm tot}$ in solar masses.
Simple self-similar scaling laws for a constant gas mass fraction
($f_{\rm gas}$) predict $L_{\rm X}\propto M_{\rm tot}$. The slight dependence
$f_{\rm gas}\propto M_{\rm tot}^{0.16}$ found by us yields $L_{\rm X}\propto M_{\rm
tot}^{1.32}$, consistent with the above result.
Schindler (1999) finds an exponent of 1.33 for a sample of high
redshift clusters using bolometric luminosities.

The good correlation shows that the X-ray luminosity is a good measure of
the mass of nearby clusters. Possible future applications include conversion of
empirical cluster luminosity functions to mass functions and
conversion of simulated mass functions to luminosity functions.
\vspace{-0.5cm} 

\end{document}